\documentclass[oldversion,letter]{aa}
\usepackage{graphicx}

\def\h2{H$_2$~}
\def\zabs{$z_{\rm abs}$~}

\begin{document}

\title{First detection of CO in a high-redshift damped Lyman-$\alpha$ system\thanks{Based on observations carried out at the European Southern
Observatory (ESO), under programme 278.A-5062 with the UVES echelle
spectrograph installed at the ESO Very Large Telescope (VLT), unit Kueyen, on
Mount Paranal in Chile.}}

\titlerunning{First detection of CO in a high-$z$ DLA}

\author{R. Srianand\inst{1}
        \and
        P. Noterdaeme\inst{2,3}
        \and
        C. Ledoux\inst{2}
        \and
        P. Petitjean\inst{3}}

\offprints{R. Srianand, anand@iucaa.ernet.in}

\institute{IUCAA, Post Bag 4, Ganeshkhind, Pune 411 007, India
          \and
          European Southern Observatory, Alonso de C\'ordova
          3107, Casilla 19001, Vitacura, Santiago 19, Chile
          \and
          UPMC Paris 06, Institut d'Astrophysique de Paris,
          UMR7095 CNRS, 98bis Boulevard Arago, F-75014, Paris, France}

\date{Received date / Accepted date}

\abstract
{We present the first detection of carbon monoxide (CO) in a damped
Lyman-$\alpha$ system (DLA) at \zabs =2.41837 toward SDSS
J143912.04$+$111740.5. We also detected \h2 and HD
molecules. The measured total column densities (in log units) of
H~{\sc i}, H$_2$, and CO are 20.10$\pm0.10$, 19.38$\pm0.10$, and
13.89$\pm$0.02, respectively. The molecular fraction, $f$ =
2$N($H$_2)$/($N$(H~{\sc i})+2$N($H$_2)$) = 0.27$^{+0.10}_{-0.08}$, is the
highest among all known DLAs. The abundances relative to solar of
S, Zn, Si, and Fe are $-$0.03$\pm$0.12, $+$0.16$\pm$0.11, $-$0.86$\pm$0.11, 
and $-$1.32$\pm$0.11, respectively,
indicating a high metal enrichment and a depletion pattern onto dust-grains
similar to the cold ISM of our Galaxy. The measured
{$N$(CO)/$N$(H$_2$)~=~3$\times$10$^{-6}$} is much less than the
conventional CO/H$_2$ ratio used to convert the CO emission into
gaseous mass but is consistent with what is measured along 
translucent sightlines in the Galaxy. The CO rotational excitation
temperatures are higher than those measured in our Galactic ISM for
similar kinetic temperature and density. Using the C~{\sc i} fine
structure absorption lines, we show that this is a consequence of the
excitation being dominated by radiative pumping by the cosmic
microwave background radiation (CMBR). From the CO excitation
temperatures, we derive $T_{\rm CMBR}$~=~9.15$\pm$0.72~K, while
9.315$\pm$0.007~K is expected from the hot big-bang theory. This is
the most precise high-redshift measurement of $T_{\rm CMBR}$ and the
first confirmation of the theory using molecular transitions at high
redshift.}

\keywords{
          Galaxies: abundances --
          Quasars: absorption lines --
          Quasars: individual: SDSS J143912.04$+$111740.5}
\maketitle


\section{Introduction}

Damped Lyman-$\alpha$ systems (DLAs) in QSO spectra are characterized by very 
high H~{\sc i} column densities, $N$(H~{\sc i})$\ga 10^{20}$ cm$^{-2}$. 
The inferred metallicities relative to solar vary between 
[Zn/H]= $-2.0$ and 0 for $2\le z_{\rm abs}\le 3$ 
(e.g. Pettini et al. 1997; Prochaska \& Wolfe 2002). 
Therefore DLAs are believed to be located in the close vicinity
of star-forming regions. 
The dust content in a typical DLA is less than or
equal to 10\% of what is seen in the Galactic 
ISM for similar $N$(H~{\sc i}), however
sufficient for favoring the formation of \h2 (Ledoux et al. 2003).

The abundance of H$_2$, the relative populations of the \h2 
rotational levels, and  the fine-structure 
levels of the  C~{\sc i} ground-state are used to derive 
the physical conditions in the gas, such as temperature,
gas pressure, and ambient radiation field (Savage et al. 
1977; Black \& van Dishoeck 1987; 
Jenkins \& Tripp 2000; Tumlinson et al. 2002). These 
conditions are  believed to be driven by the injection of 
energy and momentum through various dynamical and radiative 
processes associated with star formation activity.
Thus, detecting \h2 in DLAs at high redshifts is an important 
step forward in understanding the evolution of normal galaxies. 
Detecting other molecules would pioneer interstellar chemistry studies at 
high redshift (see e.g. Wiklind \& Combes 1995 for the dense 
ISM component).

In the course of our recently completed Very Large Telescope 
survey for H$_2$ in DLAs,  
we gathered a sample of 13 H$_2$ absorption systems at
$1.8<z<4.3$ out of a total of 77 DLAs (Ledoux et al. 2003 \& 2006; 
Petitjean et al. 2006; Noterdaeme et al. 2008). Absorption lines of
HD are detected in one of the DLAs (Varshalovich
et al. 2001) and none show detectable CO absorption.
We noticed a strong preference for H$_2$-bearing DLAs being associated 
with C~{\sc i} absorption (Srianand et al. 2005) and having high 
metallicities and  large 
depletion factors (Petitjean et al. 2006; Noterdaeme 
et al. 2008). 
In the Sloan Digital Sky Survey data base, 
we identified a most promising candidate at $z_{\rm abs}$ = 2.4185 
towards SDSS J143912.04+111740.5
showing such characteristics.
We were allocated 
8~hours of director discretionary time 
on the Ultraviolet and Visual Echelle Spectrograph (UVES) 
at the VLT of the 
European Southern Observatory (ESO)
to search for CO in addition to H$_2$. 
This observation resulted in the detection of
CO UV absorption lines that have been elusive for more than a quarter
century 
(Varshalovich \& Levshakov 1981; Srianand \& Petitjean 1998; 
Cui et al. 2005). 
We also detect H$_2$ and HD absorption lines. 
In this letter we focus our attention on the
CO excitation. A detailed analysis of the HD absorptions will be
presented elsewhere.

\section{Observations}
 
Both UVES spectrographic arms were used with standard dichroic settings and
central wavelengths of 390~nm and 580~nm (or 610~nm) for the blue and red
arms, respectively. The resulting wavelength coverage was 330$-$710~nm with a
small gap between 452 and 478~nm. The CCD pixels were binned 
$2\times 2$ and the
slit width adjusted to $1\arcsec$, yielding a resolving power
of $R=45000$ under seeing conditions of $\sim 0\farcs 9$. The total exposure
time on source exceeded 8~h.
The data were reduced using the UVES pipeline version 3.3.1 based on the
ESO common pipeline library system (M\o ller Larsen et al. 2007). Wavelengths
were rebinned to the heliocentric rest frame and individual scientific
exposures co-added. In the following, we adopt the solar reference
abundances from Morton (2003).

\section{Analysis}
\begin{figure}
\centering
\includegraphics[bb=33 160 567 693,clip,width=8.5cm,height=7.5cm, angle=0]{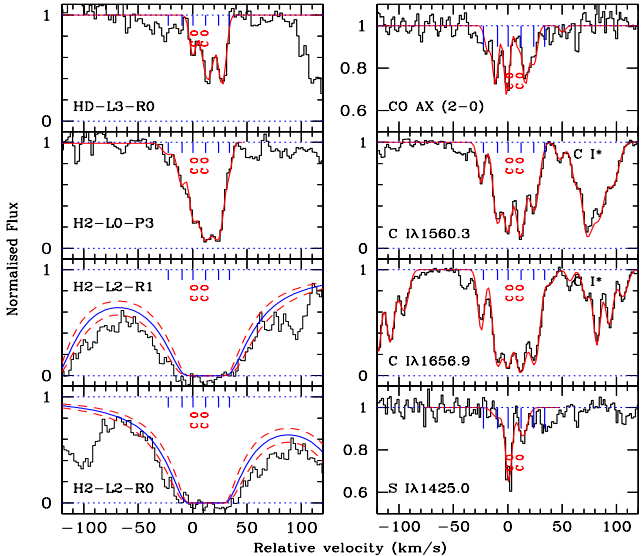}
\caption{A sample of molecular and heavy element absorption lines associated
with the damped Lyman-$\alpha$ system toward SDSS~J143912.04+111740.5. The normalized flux 
is given on a velocity scale with the origin at $z_{\rm abs} = 2.41837$. Smooth curves
in each panel show our Voigt profile fits to the data. The dashed profiles shown for
J$\le$ 1 H$_2$ lines are the 1$\sigma$ ranges. Tick marks in each panel indicate the locations of
the 6 components that are used to fit the J~=~3 H$_2$ absorption lines. 
The symbol ``CO'' marks the locations of two components detected in S~{\sc i} and CO.
}
\label{fig1}
\end{figure}

An asymmetric Lyman-$\alpha$ absorption line, together with
the corresponding Lyman-$\beta$ and $\gamma$ absorption lines, 
indicates the presence of multiple components.
The simultaneous fit to the three H~{\sc i} absorption features gives log~$N$(H~{\sc i})
(cm$^{-2}$)~=~17.8, 19.25, 19.20, 
19.40, and 20.10$\pm$0.10 at the velocities of $v$ = 
$-$892, $-$594, $-$382, $-$117, and 0 km~s$^{-1}$ relative to 
$z_{\rm abs}$ = 2.41837. The component with maximum H~{\sc i} column density 
(at $v$~=~0 km~s$^{-1}$) also exhibits H$_2$, HD, and CO absorption lines 
spread over $\sim$50 km~s$^{-1}$ (see Fig.~\ref{fig1}). 
Transitions from the J~=~3 H$_2$ rotational level are detected
in six distinct components. Both HD and CO are detected in 
three and two of these components, respectively. Transitions
from J~=~0 and 1 H$_2$ rotational levels are highly saturated,
but accurate integrated column densities can be derived from
damped wings.

Absorption lines from N~{\sc i}, O~{\sc i}, C~{\sc i}, C~{\sc i}$^{*}$, 
C~{\sc i}$^{**}$, Mg~{\sc i}, Ar~{\sc i},  S~{\sc i}, S~{\sc ii}, Si~{\sc ii}, 
Fe~{\sc ii}, Zn~{\sc ii}, Al~{\sc ii}, and Ni~{\sc ii} are seen spread over 
up to 950 km~s$^{-1}$. The total gaseous abundances 
relative to solar are 
 $-$0.03$\pm$0.12, $+$0.16$\pm$0.11, $-$0.86$\pm$0.11, 
and $-$1.32$\pm$0.11 for 
S, Zn, Si, and Fe, respectively.  
The abundances of S and Zn are consistent
with the gas abundance being close to solar, while the relative 
depletions of Si
and Fe are similar to those in cold gas in the diffuse 
Galactic ISM.
Absorption lines from the three C~{\sc i} ground-state fine-structure levels 
are detected in numerous transitions in the main H~{\sc i} component.

In Fig.~\ref{fig1} we show a few transitions from H$_2$ 
(J~=~0, 1 and 3), S~{\sc i}, 
C~{\sc i}, HD and CO. By fitting the damping wings of J~=~0  and 1 
transitions, we measured log~$N$(H$_2$, J=0)~=~18.90$\pm$0.10 and  
log~$N$(H$_2$, J=1)~=~19.18$\pm$0.10. 
We derived an excitation temperature of $T_{01}$~=~105$^{+42}_{-32}$~K, 
which is usually a good indicator of the average kinetic temperature of the gas.
The mean molecular fraction of the gas is 
$f$ = 2$N$(H$_2$)/[2$N$(H$_2$)+$N$(H~{\sc i})] = 0.27$^{+0.10}_{-0.08}$. 
This is the highest
value measured to date in a high-$z$ DLA.

\subsection{The CO molecules and the CO/\h2 ratio at high $z$}

\begin{figure}
\centering
\includegraphics[bb= 41 32 581 764,height=9.cm,width=7.5cm,clip,angle=270]{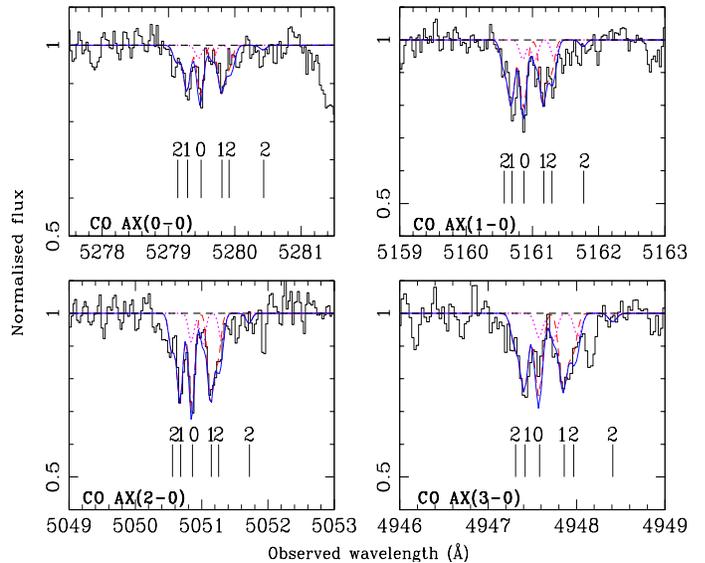}
\caption[]{Voigt profile fits to \element[][12][]CO A-X bands detected at $z_{\rm abs} = 2.4185$
towards SDSS J143912.04$+$111740.5. 
The vertical lines mark the locations of CO absorptions 
from different J levels 
at the redshift of the main component. The profiles in dashed and dotted
lines are, 
respectively, for the first and second components (at \zabs = 2.41837 and  
2.41847) and the total profile is shown in solid line.  
}
\label{fig2}
\end{figure}

{Carbon monoxide absorption 
was detected in several bands (see Fig.~\ref{fig2}). 
We used the four bands that are redshifted outside the 
Lyman-$\alpha$ forest to derive CO column densities.
Absorptions from different CO rotational levels are located 
close to each other, but the resolution of the data is high 
enough to deblend them. 
A single-component Voigt profile fit gives $z_{\rm abs}$ = 2.41837
for the main CO component. This is coincident within error ($<$1~km~s$^{-1}$)
with the strongest S~{\sc i} component (See Fig.~\ref{fig2}). 
A weaker S~{\sc i} satellite component is present at 
$z_{\rm abs}$ = 2.41853. We therefore added a second CO component with a redshift 
fixed to the value of the second S~{\sc i} component.
}
Doppler parameters were left free to vary and the best fit was obtained for $b$~=~1.5~km~s$^{-1}$, but 
the column density value in the main component is not very sensitive to the exact value for $b>0.5$~km~s$^{-1}$. 
Results are shown in Fig.~\ref{fig2}. Column densities are 
log $N$(CO)~ =~13.27$\pm$0.03, 
13.48$\pm$0.02, and 13.18$\pm$0.06, respectively, for J~=~0, 1 and 2 
in the main component. Column densities for J~=~0 and 1 are 
12.75$\pm$0.05 and 12.89$\pm$0.08 
respectively for the second component when $b$ = 1.5 km s$^{-1}$. 

We measured  $N$(CO)/$N$(H$_2$)~=~3$\times10^{-6}$.
This is similar to or slightly higher than what is 
measured along the
Galactic sightlines with similar molecular fraction and
along Galactic sightlines with  similar $N$(\h2)
(see Figs. 4 and 5 of Burgh et al. 2007). 
This is  much less than the CO/H$_2$ ratio of about 10$^{-4}$ 
derived for dense molecular clouds (e.g Lacy et al. 1994).
This strongly suggests that the physical conditions in the gas
are similar to those in the diffuse Galactic ISM.

\subsection{CO rotational excitation}

\begin{figure}
\centering
\includegraphics[bb= 22 150 576 700,width=8cm,height=7.cm,clip, angle=0]{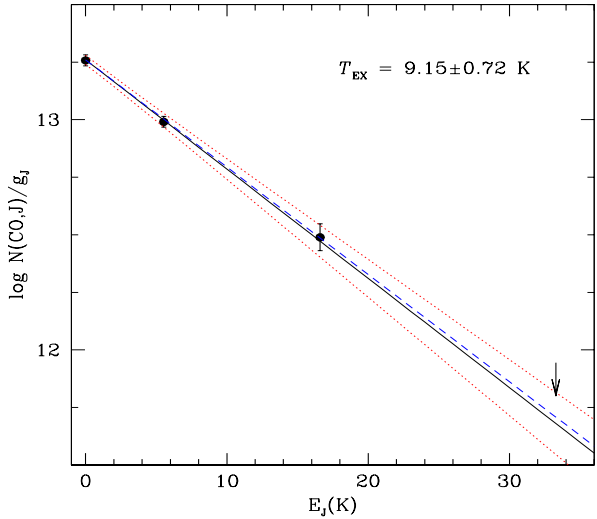}
\caption{The CO excitation diagram.
A straight line with slope 1/($T_{\rm ex}$ ln 10) indicates 
thermalization of the levels. The diagram is given for the 
main CO component at $z_{\rm abs}=2.41837$. 
The three lines give the mean and 1$\sigma$ range obtained 
from $T_{01}$,  $T_{02}$, and $T_{12}$.
The diagram is compatible with thermalization by a 
black-body radiation of
temperature 9.15$\pm$0.72~K when $T_{\rm CMBR}$ = 
9.315$\pm$0.007 K  (long dashed line)
is expected at $z_{\rm abs} = 2.4185$ from 
the hot big-bang theory.
}
\label{fig3}
\end{figure}
\begin{figure}
\centering
\includegraphics[bb=22 150 576 700,width=8.cm,height=7.cm,clip,angle=0]{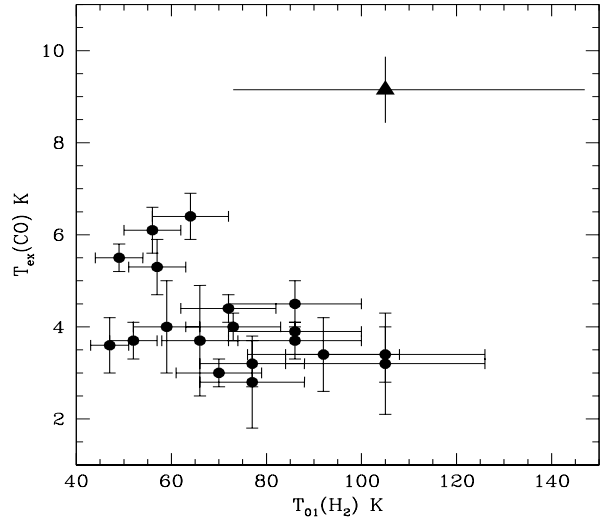}
\caption{Comparison of rotational excitation temperatures
of CO and \h2.
Filled circles are for the
measurements from Galactic diffuse ISM (Burgh et al. 2007). 
The filled triangle is for our measurement at $z = 2.4183$
towards SDSS~J143912.04+111740.5.
}
\label{fig5}
\end{figure}

The excitation temperatures derived from the population ratios of the different
rotational levels are $T_{01}=9.11\pm1.23$ K, 
$T_{12}=9.19\pm1.21$ K, and $T_{02} = 9.16\pm0.77$ K for the
main component where the errors come from the fitting uncertainties.
Additional rms deviations come from uncertainties in the continuum
placement and from the allowed range for the Doppler parameter of the
second component. We estimate these to be $\sim$0.21, $\sim$0.37, and 
$\sim$0.18 K around the three excitation temperatures.
The populations of the three rotational levels are thus consistent with 
a single excitation temperature, $T_{\rm ex} = 9.15\pm0.72$ K (see Fig. 3),
suggesting that a single mechanism controls the level populations. 
 
This excitation temperature is more than twice higher than the excitation
temperature measured in the diffuse Galactic ISM (see Fig.~\ref{fig5}), 
at similar 
H$_2$ $T_{01}$ (Burgh et al. 2007). Since we have seen that the physical 
conditions in the DLA are similar than in the diffuse Galactic ISM,
we can suspect
that the process responsible for this high excitation temperature
is specific to the high redshift of the system.
Interestingly, the population ratios in the second component are also consistent 
with this  value of $T_{\rm ex}$, albeit with large errors.

The presence of absorptions from the three fine-structure levels of the 
C~{\sc i} ground state allows us to probe the physical state 
of the gas further (Srianand et al. 2000). 
By simultaneous Voigt profile fitting of the absorption lines, we derived 
log~$N$(C~{\sc i}) = 14.26$\pm$0.01, 
log~$N$(C~{\sc i}$^*$) = 14.02$\pm$0.02, 
and log~$N$(C~{\sc i}$^{**}$) = 13.10$\pm$0.02 for the 
velocity component that corresponds to the strongest CO component. 
Assuming no contribution from the CMBR, we obtain a conservative 
hydrogen density range 
of 87-135 cm$^{-3}$ and 52-84 cm$^{-3}$
from the population ratios $N$(C~{\sc i}$^*$)/$N$(C~{\sc i}) 
and $N$(C~{\sc i}$^{**}$)/$N$(C~{\sc i}),
respectively. 
A more realistic range, 45-62 cm$^{-3}$, is obtained if we use appropriate 
excitation by the CMBR with a temperature expected from the hot big-bang
theory.
From the observed $f$ value, we derived a
conservative ($< 25$ cm$^{-3}$) and a realistic
($< 12$ cm$^{-3}$) value for the H$_2$ density 
(${n_{\rm H_2}}$~=~$f n_{\rm HI}/(2-f))$.
The above values are strict upper limits, because we have ignored the 
contribution of UV photons from stars in this galaxy 
to the C~{\sc i} excitation.             

We ran the statistical equilibrium radiative transfer code RADEX, 
available  on line (van der Tak et al. 2007), and found that 
for the kinetic temperature, $T = 105$ K, derived from H$_2$,
the  collisional contributions to $T_{01}$ and  $T_{12}$ are $\le$5\% 
and $\le 2\%$ for $n_{\rm H_2}\le 25$ cm$^{-3}$.
The corresponding values are $\le 3$\% and $\le$1\% 
for $n_{\rm H_2}\le 12$ cm$^{-3}$. Thus the collisional excitation 
of CO by H$_2$ is negligible. Collisions with H contribute little
to the CO excitation compared to collisions with H$_2$  in astrophysical conditions
(Green \& Thaddeus, 1976; Shepler et al. 2007).
This means that the CO excitation is dominated by 
CMBR, so we conclude that $T_{\rm ex}$ = $T_{\rm CMBR}$ =
9.15$\pm$0.72K. 

\section{Conclusion}
\begin{figure}
\centering
\includegraphics[bb=22 150 576 700,width=8.cm,height=7.cm,clip, angle=0]{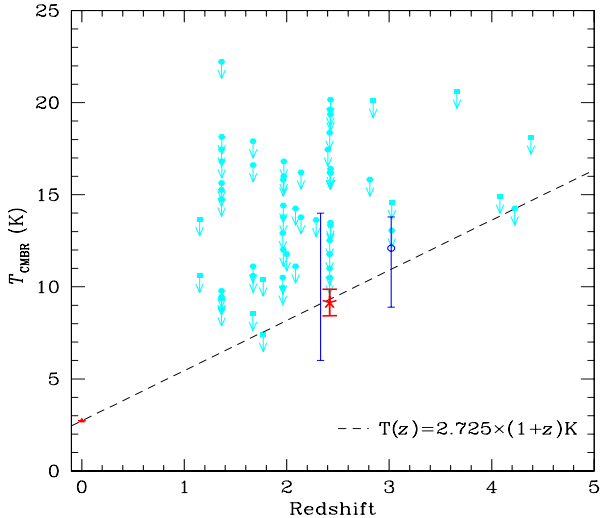}
\caption{Measurements of $T$(CMBR) at various $z$. 
The star with errorbars is for the measurement based on
CO presented here. Our earlier measurement using
fine-structure lines of neutral carbon,
$6.0<T_{\rm CMBR}<14.0$ K, at z = 2.33771 is indicated
by the long vertical bar (Srianand et al. 2000). Measurement
by Molaro et al. (2002) is marked by an open circle. 
Upper limits are measurements
using C~{\sc i} or C~{\sc ii}$^*$ 
from the literature (squares) and using C~{\sc i} 
from our UVES sample (hexagons). The dashed line is the 
prediction from the hot
big-bang theory, $T_{\rm CMBR}$($z$)=$T_{\rm CMBR}(z=0)\times(1+z)$.
$T_{\rm CMBR}$ measurement at $z=0$ is based on the 
COBE determination (Mather et al. 1999).
}
\label{fig4}
\end{figure}
{The CMBR is an important source of excitation for those species 
with transitions in the sub-millimeter range. This is the 
case for atomic species whose ground state splits into 
several fine-structure levels and of molecules that can
be excited in their rotational levels. If the relative 
level populations are thermalized by the CMBR, then the 
excitation temperature gives the temperature of the black-body 
radiation. It has long been proposed to measure the relative 
populations of such atomic levels in quasar absorption lines 
to derive $T_{\rm CMBR}$ at high redshift (Bahcall \& Wolf 1968). 
In Fig.~\ref{fig4} we combine our precise measurement of~$T_{\rm CMBR}$,
the 51 new upper limits obtained using C~{\sc i}
and C~{\sc i}$^{*}$ absorption lines detected towards QSOs in 
our UVES sample (Srianand et al. 2005; Noterdaeme et al.
2007a, 2007b; Ledoux et al. 2006), and  measurements 
reported from the literature
(Meyer et al. 1986; Songaila et al.
1994; Lu et al. 1996; Ge et al. 1997; Roth \& Bauer, 1999; Molaro
et al. 2002, Cui et al. 2005). 
Upper limits are obtained assuming CMBR as the only source of excitation.
Our precise measurements using CO and the new upper limits using C~{\sc i} 
are consistent with the adiabatic evolution of $T_{\rm CMBR}$ expected 
in the standard big-bang model (Fig.~\ref{fig4}).}

The CN molecule has proven to be a remarkable thermometer of the CMBR in 
our Galaxy.
It has been used for precise measurement of $T_{\rm CMBR}$ in different 
directions (Meyer et al. 1985; Kaiser et al. 1990). 
{Wiklind \& Combes (1996) obtained $T_{\rm CMBR}<6$ K at $z=0.885$
using the absorption lines of CS, H$^{13}$CO$^+$, and N$_2$H$^+$.}
Carbon monoxide in diffuse gas 
provides an interesting possibility for measurements at high redshift,
as the rotational energies between different rotational  
levels are close to $T_{\rm CMBR}$ at $z\ge 1$.
\par
Following careful selection of the target, based on intensive 
observations at ESO-VLT, we have made the first detection of
carbon monoxyde molecules in the diffuse ISM at high redshift.
The analysis presented here pioneers interstellar chemistry 
studies at high redshift and demonstrates that, 
together with the detection of other molecules
such as HD or CH, it will be possible to tackle important cosmological issues.
\par\bigskip\noindent
%
%
\begin{acknowledgements}
We warmly thank the Director
Discretionary Time allocation committee and the ESO Director General,
Catherine Cesarsky, for allowing us to carry out these observations.
RS and PPJ gratefully acknowledge support from the Indo-French Center
for the Promotion of Advanced Research (Centre Franco-Indien pour la
Promotion de la Recherche Avanc\'ee) under contract No. 3004-3.  PN is
supported by an ESO PhD studentship. 
\end{acknowledgements}


\end{document}